\documentclass[12pt,preprint]{aastex}
\newcommand{\be}{\begin{equation}}
\newcommand{\ee}{\end{equation}}
\newcommand{\bea}{\begin{eqnarray}}
\newcommand{\eea}{\end{eqnarray}}
\shorttitle{New Regime of MHD Turbulence}
\begin{document}
\title{New Regime of MHD Turbulence: Cascade Below Viscous Cutoff}
\author{Jungyeon Cho \& Alex Lazarian}
\affil{Dept. of Astronomy, University of Wisconsin,
    475 N. Charter St., Madison, WI53706; cho, lazarian@astro.wisc.edu}

\author{Ethan T. Vishniac}
\affil{Dept. of Physics and Astronomy, Johns Hopkins University,
3400 N. Charles St., Baltimore, MD 21218; ethan@pha.jhu.edu}

\begin{abstract}
%version 1.1: (05-15-01) 11-27-01 to 11-27-01
In astrophysical situations, e.g. in the interstellar medium (ISM),
neutrals can provide viscous damping on scales 
much larger than the magnetic diffusion scale.
Through numerical simulations,
we have found that the magnetic field can have a rich structure below
the dissipation cutoff scale.
This implies that magnetic fields in the ISM can have structures
on scales much smaller than parsec scales. 
Our results show that the magnetic energy contained in a 
wavenumber
band is independent of the wavenumber 
and magnetic structures are intermittent and extremely anisotropic.
We discuss the relation between our results and the formation of
the tiny-scale atomic structure (TSAS).

\end{abstract}

%%%%%%%%%%%%%%%%%%%%%%%%%%%%%%%%%%%%%%%%%%%%%%%%%%%%%%%%%%%%%%%%%%%%%

\section{Introduction}
In the interstellar medium (ISM),
flows are complicated and dynamic.
Observations suggest that the ISM is in a turbulent state
(Larson 1981; Myers 1983; Scalo 1984; Armstrong, Rickett \& Spangler 1995;
Stanimirovic \& Lazarian 2001).

Hydrodynamic turbulence can be described by so-called
energy cascade model, in which energy injected at a
scale $L$ cascades down to progressively smaller and smaller scales.
Ultimately, the energy will reach the molecular 
{\it dissipation scale} $l_d$
and the energy will be lost there.
The scales between $L$ and $l_d$ constitute the {\it inertial range}.
In hydrodynamic turbulence the dissipation scale is the 
minimal scale for motion. If we plot kinetic energy spectrum $E_v(k)$, 
the kinetic energy contained in a 
wavenumber band, we will see that
a) the spectrum peaks at the wavenumber corresponding to the
energy injection scale ($k_L \sim 1/L$); 
b) it follows a power law (e.g. $E_v(k) \propto k^{-5/3}$ in Kolmogorov
theory) in the inertial range; 
c) it drops rapidly after $k_d \sim 1/l_d$, which depends on
viscosity $\nu$: $k_d \sim (VL/\nu)^{3/4}k_L$, where $V$ is the rms velocity
at the energy injection scale $L$.

Magnetohydrodynamic (MHD) turbulence  has two energy loss scales -
a viscous damping scale set by the viscosity $\nu$ and magnetic 
diffusion scale set by the ohmic resistivity $\eta$.
When $\nu\gg\eta$,
the viscous damping scale is much larger than 
the magnetic diffusion scale.
Although MHD turbulence is different from its hydrodynamic counterpart
in many ways,
the energy cascade model is still valid (e.g. 
%%%Iroshnikov 1963; Kreichnan 1965; 
Goldreich \& Sridhar 1995).
Therefore, in this case, kinetic energy is damped before the cascading energy reaches
the magnetic diffusion scale. In this paper we study MHD
turbulence when the mean field $B_0$ is at least as large as 
the fluctuating field $b$. This is the opposite of the 
dynamo regime with $B_0\ll b$ discussed in
Kulsrud \& Anderson (1992) and tested in Maron \& Cowley (2001).
We shall show that the 
wide-spread assumption that magnetic structures do not exist
below the viscous cutoff is wrong. 

In astrophysics, the viscosity caused by neutrals damps turbulence.
In the ISM, this viscous cutoff occurs 
at $\sim$~$pc$ scales,
which is much larger than the magnetic diffusion scale.
To model this, we use a large physical viscosity and very small
magnetic diffusivity.

In this letter we numerically demonstrate the existence of a power-law
magnetic energy spectrum below the viscous damping scale.
We use an incompressible MHD code.
%%%pseudo-spectral incompressible MHD code.

%%%%%%%%%%%%%%%%%%%%%%%%%%%%%%%%%%%%%%%%%%%%%%%%%%%%%%%%%%%%%%%%%%%%%

%%%%%%%%%%%%%%%%%%%%%%%%%%%%%%%%%%%%%%%%%%%%%%%%%%%%%%%%%%%%%%%%%%%%%

\section{Method}
\noindent {\bf Numerical Method.}
We have calculated the time evolution of incompressible magnetic turbulence
subject to a random driving force per unit mass.
We have adopted a pseudo-spectral code to solve the
incompressible MHD equations in a periodic box of size $2\pi$
(see more in Lazarian, Vishniac \& Cho 2002):
\begin{equation}
\frac{\partial {\bf v} }{\partial t} = -(\nabla \times {\bf v}) \times {\bf v}
      +(\nabla \times {\bf B})
        \times {\bf B} + \nu \nabla^{2} {\bf v} + {\bf f} + \nabla P' ,
        \label{veq}
\end{equation}
\begin{equation}
\frac{\partial {\bf B}}{\partial t}=
     \nabla \times ({\bf v} \times{\bf B}) + \eta \nabla^{2} {\bf B} ,
     \label{beq}
\end{equation}
%\be
%      \nabla \cdot {\bf v} =\nabla \cdot {\bf B}= 0,
%\ee
where $\bf{f}$ is a random driving force,
$P'\equiv P/\rho + v^2/2$, ${\bf v}$ is the velocity,
and ${\bf B}$ is magnetic field divided by $(4\pi \rho)^{1/2}$.
In this representation, ${\bf v}$ can be viewed as the velocity 
measured in units of the rms velocity
of the system and ${\bf B}$ as the Alfv\'en speed in the same units.
The time $t$ is in units of the large eddy turnover time ($\sim L/V$) and
the length in units of $L$, the scale of the energy injection.
In this system of units, the viscosity $\nu$ and magnetic diffusivity $\eta$
are the inverse of the kinetic and magnetic Reynolds numbers respectively.
The magnetic field consists of the uniform background field and a
fluctuating field: ${\bf B}= {\bf B}_0 + {\bf b}$.
The Alfv\'en speed of
the background field, $B_0$, is set to 1.
We use the same numerical technique as in Cho \& Vishniac (2000).

We use a physical viscosity ($\nu=0.015$) and
hyper diffusion.
The power of hyperdiffusion
is set to 2 or 3, so that the magnetic dissipation term in the above equation
is replaced with
$
 -\eta_n (\nabla^2)^n {\bf v},
$
where n = 2 or 3 and 
$\eta_n$ (=$\eta_2$ or $\eta_3$ ) is adjusted in such a way that
the magnetic dissipation occurs around $k\sim N/3$, 
where $N$ is the number of grids in each spatial direction. 
This way, we can avoid the aliasing error
of the pseudo-spectral method.
%%%We list parameters used for the simulations in Table 1.
%%%Diagnostics for our code can be found in Cho and Vishniac (2000a).
We use the notation 384XY-$B_0$Z,
where 384 refers to the number of grid points in each spatial
direction; X=P refers to physical viscosity; 
Y = H2, H3 refers to hyper-diffusion 
(and its power);
Z=1 refers to the strength of the external magnetic field.

\vspace{1 mm}
\noindent {\bf Parameter Space.}
We require that the Alfv\'en frequency is similar to the eddy turnover rate
at the energy injection scale: $B_0 k_{\|,L} \approx V k_{\perp, L}$,
where $k_{\|,L}$ and $k_{\perp,L}$ are the wavenumbers parallel and 
perpendicular 
to 
${\bf B}_0$ at the scale, respectively.
%%%$B_0 \approx V$.
We require a large viscosity to maximize the dynamical range of the
damped, conducting regime, but we need $\nu$ small enough to guarantee 
large scale turbulence.
Therefore, we take $\nu \sim 0.015$ (for a Reynolds number 
$R \equiv LV/\nu \sim 100$).
When we take 
$\nu=0.015$, $k_L \sim 2.5$, and $V \sim 0.6$,
then $k_d \equiv R^{3/4} k_L \sim 70$. 
In practice, the kinetic energy
spectrum usually begins to drops rapidly 
one decade before $k_d$, so the
viscous cutoff scale is around $k=7$.
We also require $\nu \gg \eta$.
Therefore, we take hyper-diffusion for magnetic field.
We consider only low order hyper-diffusion.
Using hyper-diffusion, we can have a very small magnetic diffusion
for small k's and very strong diffusion at large k's so that
the large scale magnetic field is largely unaffected by ohmic dissipation.

%%%%%%%%%%%%%%%%%%%%%%%%%%%%%%%%%%%%%%%%%%%%%%%%%%%%%%%%%%%%%%%%%%%%%
\section{Results}

\begin{figure}[h!t]
%{\centering  \leavevmode 
%\epsfxsize=3.2in \epsfbox{f1.eps}
%\hfil 
%\epsfxsize=3.0in \epsfbox{f2.eps}} 
\plottwo{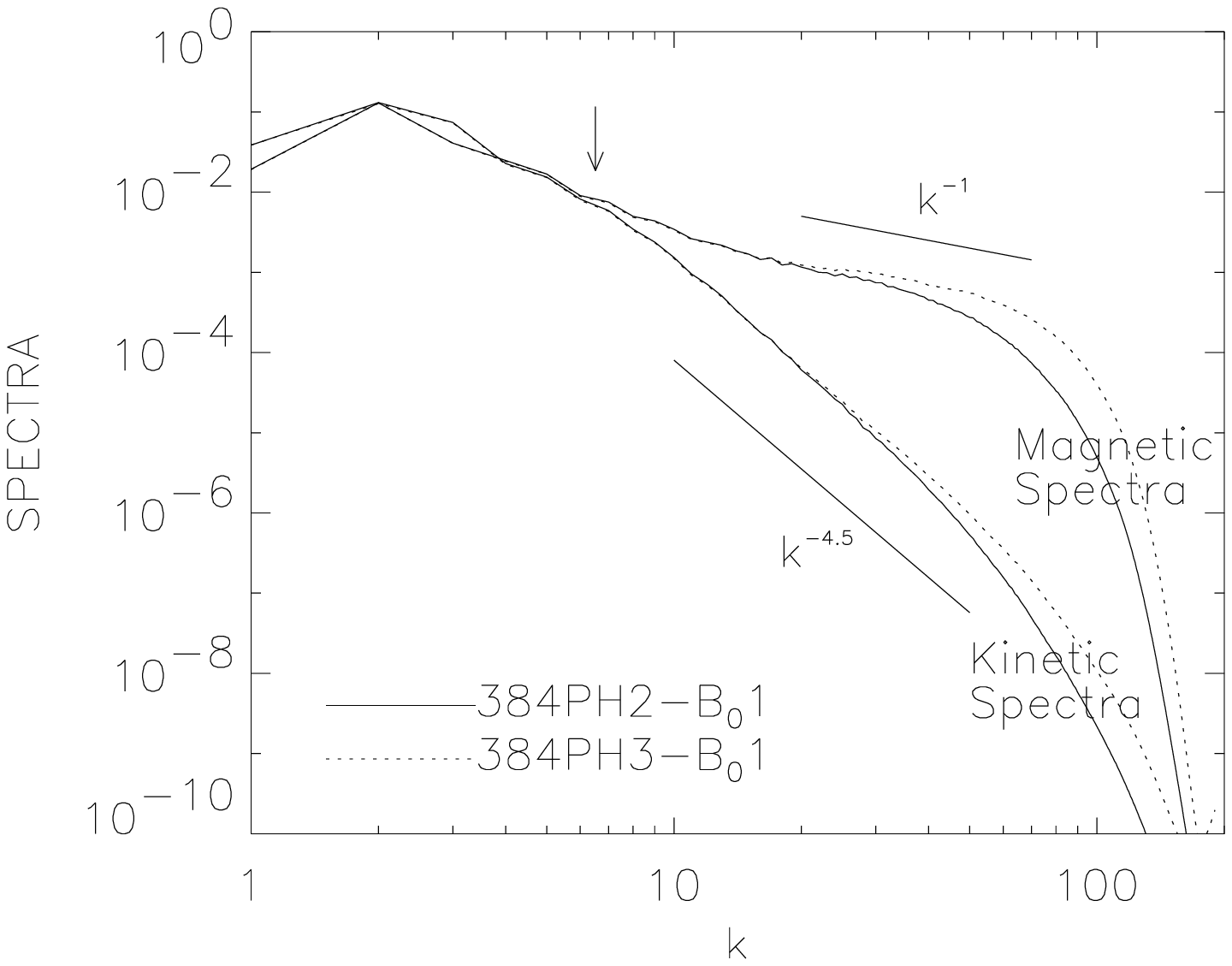}{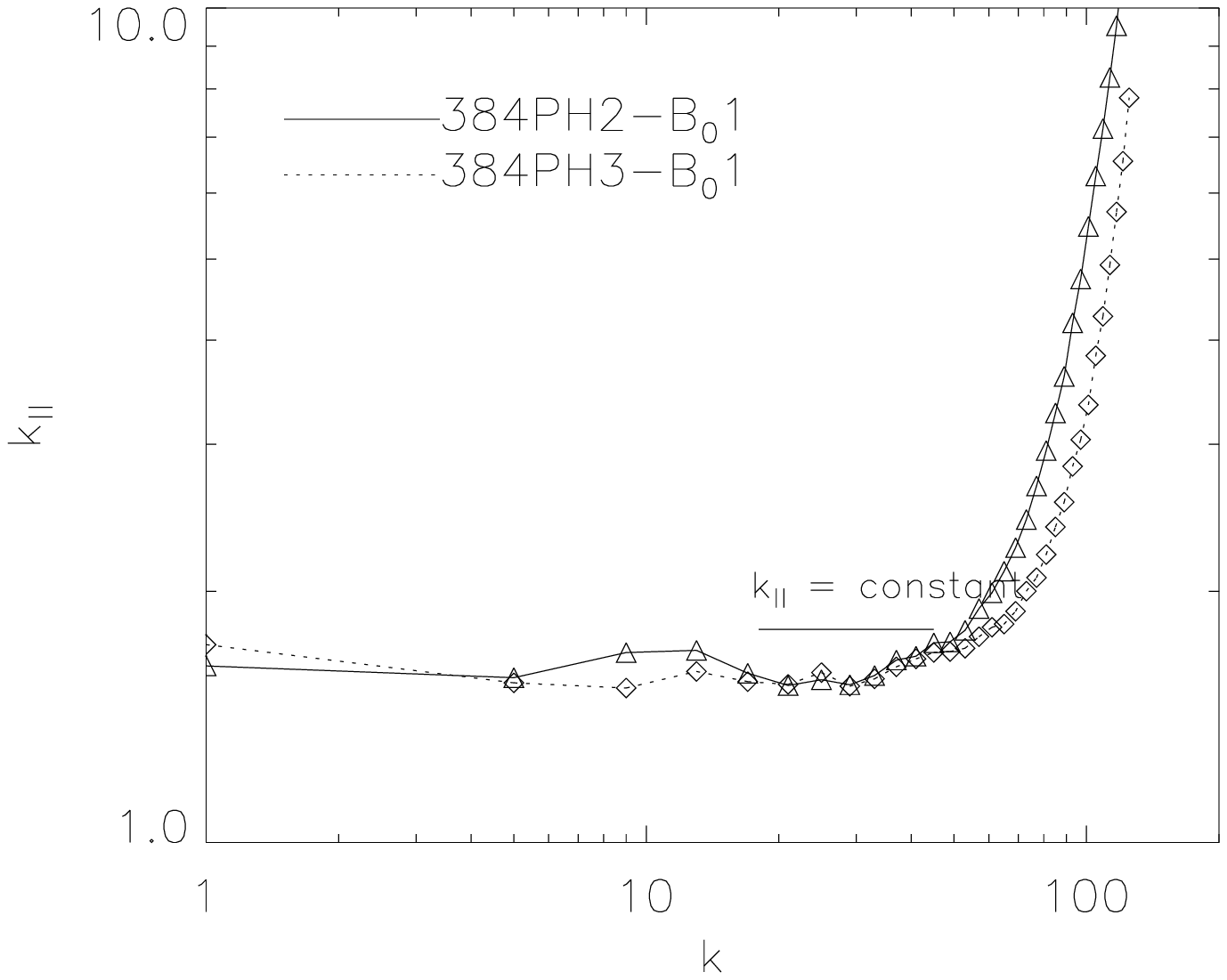}
\caption{ 
   Energy spectra $E_v(k)$ and $E_b(k)$ at t=12.
   Spectra of 384PH2-B$_0$1 and 384PH3-B$_0$1 almost exactly coincide
   at small wavenumbers. Run 384PH3-B$_0$1 has more extended inertial range
   below the viscous cutoff (marked by an arrow).
   }
\caption{ 
   Parallel wavenumber $k_{\|}$. Solid line: 384PH2-B$_0$1 at t=10.
   Dotted line: 384PH3-B$_0$1 at t=12.
   }
\end{figure}

\noindent {\bf Time Evolution and Spectra.}
About 4 to 5 time units after the start of the simulation, 
the system
has reached a statistically stationary state, where
the average rms velocity ($\sim$~$0.6$) 
is roughly equal to the average strength
of the random magnetic field.
Energy spectra reach a statistically stationary state a bit later
at $\sim$~$6$.
{}For the run 384PH2-B$_0$1 we integrate the MHD 
equations from $t=0$ to $t=12$.
{}For the run 384PH3-B$_0$1 we integrate the MHD 
equations from $t=8$ to $t=12$, where we use a data cube from the run
384PH2-B$_0$1 at $t=8$ as the input.
We calculate energy spectra and other quantities at $t\sim$~$12$.

In Figure 1, we plot energy spectra at t=12.
The spectra consist of several parts.
{}First, the peak of the spectra corresponds to the energy injection scale.
Second, for $2<k<7$, kinetic and magnetic spectra follow a similar slope.
This part is more or less a severely truncated inertial range for undamped MHD
turbulence.  Third, the magnetic and kinetic spectra begin to decouple at $k\sim 7$.
Fourth, after $k\sim20$, a new damped-scale inertial 
range emerges.
%%%
We checked that the tail of the magnetic fluctuations is a real physical effect
and not due to a bottle-neck, by comparing with 
calculations with a real magnetic diffusivity
and obtaining a similar effect.
%%%%%%%%%%%%%%%%%%%%%%%%%%%%%%%%%%%%%%%%%%%%%%%%%%%%%%%%%%%%%%%%%
%%Although a high order hyperdiffusion simulation 
%may suffer the bottle-neck effect,
%there is no evidence that low order one (e.g. 2nd or 3rd order)
%is affected by the effect.
%We also have evidence that the new inertial range in Figure 2 is not 
%by the bottle-neck effect.
%First, according to Cho \& Vishniac (2000) the bottle-neck effect in MHD
%is not as prominent as in the hydro cases.
%Second, the bottle-neck effect changes the slope of the energy spectrum
%right before the dissipation cutoff. Usually it affects Fourier modes
%in the range of $k_d/2 < k < k_d$, 
%where $k_d\sim 130$ is the magnetic dissipation wave number.
%%We can see that the new inertial range emerges before $k_d/2$.
%%%%%%%%%%%%%%%%%%%%%%%%%%%%%%%%%%%%%%%%%%%%%%%%%%%%%%%%%%%%%%%%%

In summary, scaling for energy spectra {\it below the viscous cutoff} 
is as follows.
The magnetic spectrum from Run 384PH2-B$_0$1 follows
\begin{equation}
  E_b(k) \propto k^{-1}.
\end{equation}
while Run 384PH3-B$_0$1 shows a slightly shallower slope.
The kinetic energy spectra are roughly
\begin{equation}
  E_v(k) \propto k^{-4.5}.
\end{equation}

\vspace{1 mm}
\noindent {\bf Scaling of $k_{\|}$.}
%%%%%%%%%%%%%%%%%%%%%%%%%%%%%%%%%%%%%%%%%%%%%%%%%%%%%%%%%%%%%%%
%When the slope of energy spectrum is steeper than $-3$, 
%one cannot use structure functions to describe turbulence.
%Therefore, it is necessary to use other methods.
%Here we introduce a new method to calculate 
%$k_{\|}$ as a function of $k_{\perp}$.
% Guys, this is a funny claim.  The magnetic field spectrum is
% not steep at all.  Perhaps you mean it's steep in terms of
% k_{\|}?  Better to say something like this:
%%%%%%%%%%%%%%%%%%%%%%%%%%%%%%%%%%%%%%%%%%%%%%%%%%%%%%%%%%%%%%%
{}For the kind of intermittent and extremely anisotropic structures
seen these simulations, the interpretation of our results in
terms of structure functions is complicated.  Here we concentrate
on $k_{\|}(k_{\perp})$ as a measure of field line curvature.
There may be other independent length scales parallel to ${\bf B}_0$ 
characterizing the properties of the magnetic field structures. 

The term ${\bf B}\cdot \nabla {\bf B}$ describes magnetic tension
and is approximately equal to $B_0 k_{\|} b_l$ for an isolated eddy
in a uniform mean field ${\bf B}_0$. 
Here $k_{\|} \propto 1/l_{\|}$ and $l_{\|}$
is the characteristic length scale parallel to ${\bf B}_0$, which
is known to be larger than the perpendicular length scale.
%%%
Cho \& Vishniac (2000) and
Cho, Lazarian \& Vishniac (2002) argued that, in actual turbulence, 
eddies are aligned with the local mean field ${\bf B}_L$.
%%%
We can obtain the {\it local frame representation} of
$k_{\|}$, by considering an eddy lying 
in the {\it local} mean field ${\bf B}_L$: 
${\bf B}_L\cdot \nabla {\bf b}_l \approx B_L k_{\|} b_l.$\footnote{
There can be many ways to define the local mean field ${\bf B}_L$.
In this paper, we obtain ${\bf B}_L$ for an eddy of (perpendicular) 
size $l\propto 1/k$
  by eliminating
the Fourier modes whose perpendicular wavenumber is greater than $k/2$ and
  ${\bf b}_l$  by eliminating
the Fourier modes  whose perpendicular wavenumber is less than $k/2$.
}
{}Fourier transform of both sides yields
%%%\be
$
 |\widehat{ {\bf B}_L\cdot \nabla {\bf b}_l }|_{\bf k}
    \approx B_L k_{\|} |\hat{b}|_{\bf k},
$
%%%\ee
where hatted variables are Fourier-transformed quantities.
{}From this, we have
\be
k_{\|} \approx \left( 
     \frac{\sum_{k\leq |{\bf k}^{\prime}| <k+1} 
   |\widehat{ {\bf B}_L\cdot \nabla {\bf b}_l   }|_{{\bf k}^{\prime}}^2  }
  { B_L^2 \sum_{k\leq |{\bf k}^{\prime}| <k+1} |\hat{b}|^2_{{\bf k}^{\prime}} }
               \right)^{1/2}.    \label{kparLl}
\ee

{}Figure 2 shows that by this measure $k_{\|}$ is nearly constant:
\begin{equation}
  k_{\|} \approx k_{\|,d} = \mbox{constant,}
\end{equation}
where $k_{\|,d}$ is the parallel wavenumber 
at the viscous damping scale.\footnote{
Constancy of $k_{\|}$ means an extreme form of anisotropy: 
eddies have a constant parallel size 
while they can have extremely small perpendicular size.}
In the second-order hyperdiffusion run 384PH2-B$_0$1, 
$k_{\|}$ is almost constant up to $k\sim 50$.
The sharp rise after $k\sim 50$ is due to magnetic dissipation and the
exponential suppression of the magnetic field energy.
The third order hyperdiffusion run (the dotted line) shows more extended
range of constant $k_{\|}$ since magnetic dissipation occurs at larger $k$'s.

We checked whether or not equation (\ref{kparLl}) 
gives a reasonable result 
for ordinary (i.e. {\it not} viscously damped) MHD turbulence.
Cho \& Vishniac (2000) 
showed numerically that
$k_{\|} \propto k_{\perp}^{2/3}$ using structure functions 
when the turbulence is threaded by
a strong external magnetic field (see Goldreich \& Sridhar 1995 for a theoretical
derivation).
When we applied the method in equation (\ref{kparLl}) to this
case we recovered the relation $k_{\|} \propto k_{\perp}^{2/3}$, confirming
our intuitive notion that equation (\ref{kparLl})
can provide us with a coordinate-independent way of
calculating $k_{\|}$ as a function of $k_{\perp}$.

\vspace{1 mm}
\noindent {\bf Intermittency.}
In Figure 3, we show strength of small scale magnetic fields 
in a plane perpendicular to the mean field ${\bf B}_0$.
Darker tones represent stronger magnetic fields.
We also plot small scale magnetic vectors projected on the plane.
We obtain the small scale magnetic field by eliminating
Fourier modes with $k<20$. Only part of the plane is shown here.
We see that the magnetic structures are very intermittent.\footnote{
This is a real physical effect. When we use a real magnetic 
diffusivity, we also obtain highly intermittent structures similar to
those in Fig.~3.}
We also see that 
the typical radius of curvature of field lines in the plane
is much larger than the typical perpendicular scale for field reversal.
The typical radius of curvature of field lines corresponds to the
viscous damping scale, indicating that stretched structures are
results of the shearing motions at the viscous damping scale.
This is consistent with Figure 2, which shows that the parallel
wavelength at the viscous damping scale is also reflected in the
curvature of the small scale field lines.
There is no preferred direction for these elongated structures.
A similar plot for ordinary ({\it not} viscously damped) MHD turbulence
shows much less intermittent structures.

\vspace{1 mm}
\noindent {\bf What is the Scaling?}
In this section, we give a simple set of scaling relations for the
properties of this regime (e.g. $20<k<50$ for Run 384PH3-B$_0$1).
Later, in Lazarian, Vishniac \& Cho (2002, henceforth LVC02), we will discuss
the physical basis for these scaling relations in detail.

We first note that the magnetic spectrum is roughly proportional to
$k^{-1}$.
When the shearing by the motion at scale $l_d$ dominates that of 
the scale $l$, the damping-scale shearing timescale ($l_d/v_d$) becomes the
characteristic timescale at the scale $l$.
Note that the characteristic timescale is scale-independent.
Assuming a local cascade of energy in phase space we have
\begin{equation}
   b_l^2/(l_d/v_d) \approx \mbox{constant,}
\end{equation}
so that
%%%\begin{equation}
$
   b_l \approx \mbox{constant,}
$
%%%\end{equation}
which, in turn, implies 
\begin{equation}
   E_b(k) \propto k^{-1}.
\end{equation}
This is identical to the viscous-convective range of a passive scalar
in hydrodynamic turbulence (see, for example, Lesieur 1990).

{}Figure 2 suggests that $k_{\|}\approx$constant below the cutoff.

Figure 3 shows that the magnetic structures are very intermittent below the 
viscous cutoff.  Assuming that the filaments are characterized by a rough
balance between viscous drag and magnetic tension, calculations in LVC02 predict
that 
\begin{equation}
   E_v(k) \propto k^{-4}
\end{equation}
which is almost exactly true in our simulations.

\section{Implications} 
The intermittent small scale structures that we predict here should have 
important implications for transport processes (heat, cosmic rays, etc.)
in partially ionized plasmas.
We also speculate that they might have some relation to 
the tiny-scale atomic structures (TSAS).
Heiles (1997) introduced the term TSAS
for the mysterious
H~I absorbing structures on the scale from thousands to tens of
AU, discovered by Deiter, Welch \& Romney (1976). Analogs are observed
in NaI and CaII (Meyer \& Blades 1996; Faison \& Goss 2001; 
Andrews, Meyer \& Lauroesch 2001) and in molecular gas 
(Marscher, Moore \& Bania 1993). 
Recently Deshpande, Dwarakanath 
\& Goss (2000) 
analyzed channel maps of opacity fluctuations toward Cas A and Cygnus A.
They found 
that the amplitudes of density fluctuations at scales less than 0.1 pc 
are far larger than expected from extrapolation from larger scales, 
possibly explaining TSAS. This study, however, cannot answer what 
confines those presumably overpressured (but very quiescent!) blobs of 
gas. Deshpande (2000) related those structures to the shallow spectrum of
interstellar turbulence.
 
{}Figure 1 indicates that while velocity decreases rapidly, but {\it not}
exponentially, below the viscous damping scale,
the magnetic field fluctuations persist, 
thereby providing nonthermal pressure. Magnetic structures 
perpendicular to the mean magnetic field are compensated by pressure 
gradients from density fluctuations reminiscent of the Dishpande et al. 
(2000) observations. 
 
Our calculations are applicable on scales from the viscous damping 
scale (determined by equating the energy transfer rate with the 
viscous damping rate; $\sim0.1$ pc in the Warm Neutral Medium with $n$ 
= 0.4 cm$^{-3}$, $T$= 6000 K) to the ion-neutral decoupling scale (the 
scale at which viscous drag on ions becomes comparable to the neutral 
drag; $\ll 0.1$ pc). Below the viscous scale the fluctuations of 
magnetic field obey the damped regime shown in Figure 1 and produce 
density fluctuations. If gas is close to be thermally unstable 
(Vazquez-Semadeni, Gazol \& Scalo 2000), 
moderate fluctuations of pressure cause substantial 
density excursions. For typical Cold Neutral Medium gas, the scale of 
neutral-ion decoupling decreases to $\sim70$AU, and is less for denser 
gas. TSAS may be created by strongly nonlinear MHD turbulence! 
 
The issue of an adequate description of turbulence in the newly found 
regime of persistent turbulence goes much beyond the TSAS 
formation. The accumulation of energy at small scales is important for 
problems of cosmic ray transport, magnetic reconnection, and formation 
of clumps in molecular clouds. We expect to make significant progress 
in those fundamentally important directions as soon as we describe 
properly the new regime of MHD turbulence.

\section{Conclusion}
We have considered MHD turbulence with $\nu \gg \eta$, which
implies that the viscous cutoff occurs at a scale much larger
than the magnetic diffusion scale.
It has been believed that
the damping of the fluid motion is accompanied by the suppression of
magnetic structures below the viscous cutoff scale.
To the contrary, we have found that the
magnetic field can have a rich structure below the viscous cutoff scale.
Judging from our (limited) numerical results, we conclude that
magnetic field perturbations have a similar power distribution as a passive scalar,
despite the obvious importance of magnetic forces.
In particular, the spectrum follow a $k^{-1}$ law.
The parallel wavenumber, which is an indicator of the degree of anisotropy,
is almost constant. Consequently,  the magnetic field has an
extreme form of anisotropy.
In summary, we have found that
\begin{enumerate}
\item $E_b(k)\propto k^{-1}$, 
\item $E_v(k)\propto k^{-4}$, 
\item $k_{\|} \approx \mbox{constant.}$
\end{enumerate}
We discussed the possibility that
this small scale magnetic field is the cause of the 
tiny-scale atomic structure.
The small scale magnetic structure will affect many astrophysical
processes (e.g. cosmic-ray transport, reconnection, etc) that 
depend on the statistical properties of MHD turbulence.

\acknowledgements
J.C. thanks Peter Goldreich for clarifying the idea of calculating $k_{\|}$.
A.L. and J.C. acknowledge the support of NSF Grant AST-0125544.
E.V. acknowledges the support of NSF Grant AST-0098615.
This work was partially supported by National Computational Science
Alliance under AST000010N and
utilized the NCSA SGI/CRAY Origin2000.

\begin{figure}[h!t]
%{\centering  \leavevmode 
%\epsfxsize=3.0in \epsfbox{f3.eps}
%\hfil} 
%\epsfxsize=3.0in \epsfbox{f4.eps}} 
\plotone{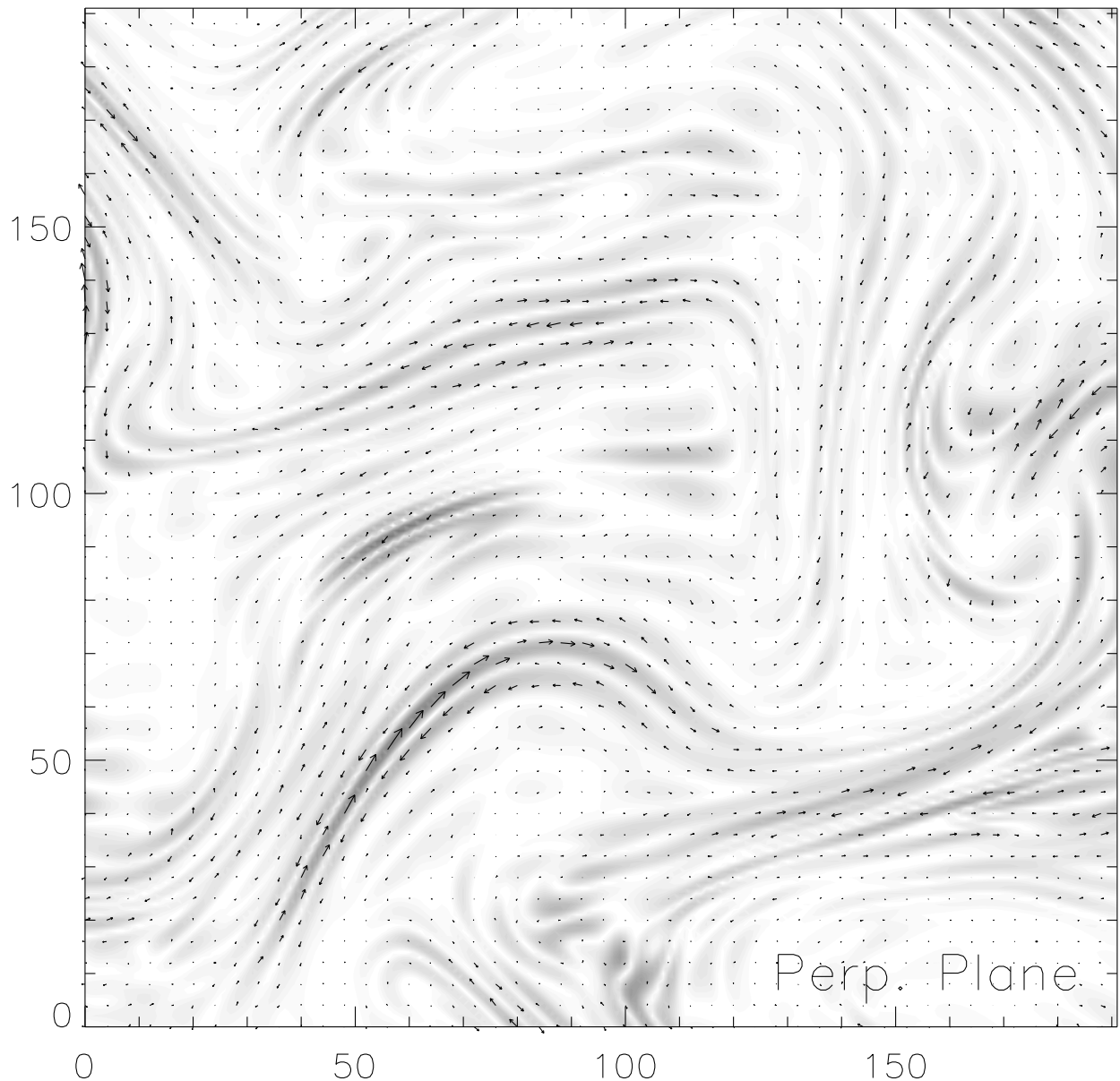}
\caption{ 
  Strength of magnetic field in a plane perpendicular to ${\bf B}_0$ at t=12.
  Arrows are magnetic fields in the plane.
  Only 1/4 is shown. Note highly intermittent structures. Run 384PH3-B$_0$1.
   }
%%\caption{ Comparison figure.
%%  Same as Figure 3, but for ordinary (not viscous-damped) MHD turbulence.
%%  Hyper-viscosity and hyper-diffusion are used.
%%  Run 256H8H8-B$_0$1, the same run as 256H-B$_0$1 
%%  in Cho \& Vishniac (2000b).
%%   }
\end{figure}

\end{document}